\documentclass{jpsj2}
\title{Gauge Theory for Quantum Spin Glasses}

\author{Satoshi \textsc{Morita}$^{1}$, Yukiyasu \textsc{Ozeki}$^{2}$
and Hidetoshi \textsc{Nishimori}$^{1}$}

\inst{$^{1}$Department of Physics, Tokyo Institute of Technology,
Oh-okayama, Meguro-ku, Tokyo 152-8551\\
$^{2}$Department of Applied Physics and Chemistry, The University of
Electro-Communications, Chofugaoka, Chofu-shi, Tokyo 182-8585}

\abst{The gauge theory for random spin systems is extended to quantum
spin glasses to derive a number of exact and/or rigorous results.  The
transverse Ising model and the quantum gauge glass are shown to be gauge
invariant. For these models, an identity is proved that the expectation
value of the gauge invariant operator in the ferromagnetic limit is
equal to the one in the classical equilibrium state on the Nishimori
line. As a result, a set of inequalities for the correlation function
are proved, which restrict the location of the ordered phase. It is also
proved that there is no long-range order in the two-dimensional
quantum gauge glass in
the ground state. The phase diagram for the quantum $XY$ Mattis model is
determined.}

\kword{spin glass, gauge theory, correlation function,
 quantum spin system}

\begin{document}
\maketitle

\section{Introduction}
The problem of spin glasses has been attracting continued
attention.\cite{BP,FH} After the pioneering work by Edwards and Anderson
(EA)\cite{EA}, Sherrington and Kirkpatrick have exactly solved the
mean-field model by assuming the replica symmetry.\cite{SK} Parisi has
proposed the replica symmetry breaking solution and established the
theoretical picture that the low-temperature phase is composed of
infinitely many stable states with ultrametric structure\cite{Parisi}.

A topic of active investigations in recent years concerns the properties
of short-range systems. Numerical studies have provided strong evidence
for the existence of the spin glass (SG) transition for both the EA
model\cite{BY,OM,SC,BY2} and the $XY$ gauge glass\cite{HS,RTYF,CBK,KS}
in three dimensions but against it for the two-dimensional EA
model\cite{SC,BY2}. For the two-dimensional gauge glass, although the
long-range SG order has been denied rigorously,\cite{NK} it
is still possible that the system has a quasi long-range order in
which the SG correlation decays in a power low. There remains the
controversy about the existence of this order: some numerical studies
have supported the absent of a finite-temperature
transition\cite{KS,G,RY} but some groups argue against such a
conclusion.\cite{L,CP}

Analytical calculations for spin glasses in finite dimensions are
difficult because of randomness and frustration. However, a method using
the gauge symmetry of the system is well-known as a powerful
technique. \cite{N_ptp,N_ox} This method provides various rigorous
results, for instance, the exact internal energy and an upper bound for
the specific heat in the special region of the phase diagram. Another
noteworthy result is a set of inequalities for the correlation function,
which restrict the topology of the phase diagram. In this relation,
it has been suggested that the phase boundary between the 
ferromagnetic (FM)
and SG phases is vertical by modifying the probability
distribution.\cite{Kita} These results are generalized to a wider class
of systems including the usual Ising SG and the $Z_q$ and $XY$
gauge glasses.\cite{ON}

Although the gauge theory provides us with surprising results, its
targets have so far been limited to classical spin systems. In the present
paper, we generalize this theory so that it applies to quantum spin
systems. A difficulty of this generalization is the fact that we must
define the gauge transformation for spins without violating the
commutation rule. We circumvent this problem by using a rotational
operator on the Hilbert space as the gauge transformation.

This paper consists of six sections. In the next section, we formulate
the gauge transformation in two quantum spin glasses, the transverse
Ising model and the quantum gauge glass (QGG), and show that these
models have gauge symmetry. In \S 3, we prove an identity for a gauge
invariant operator. This identity is valid even when the system
parameters depend on time following the time-dependent Schr\"{o}dinger
equation. In \S 4, we derive a set of inequalities for correlation
functions and order parameters. These results restrict the location of
the FM phase or the Kosterlitz-Thouless (KT) phase in the phase
diagram. In \S 5, we extend these inequalities
to the ground state. The resulting inequalities for
the order parameters show that the FM order does not exist at
zero-temperature in the two-dimensional QGG.
In \S 6, we consider the quantum $XY$ Mattis model and
determine its phase diagram. The last section is devoted to summary.

\section{Gauge Transformation for Random Quantum Spin Systems}

\subsection{Transverse Ising model}
First, let us consider the random-bond Ising model in a transverse
field. The Hamiltonian for this model is written as
\begin{equation}
 H=-\sum_{\langle ij \rangle} J_{ij}\sigma_i^z \sigma_j^z
  - h \sum_{i} \sigma_i^x, \label{TIM_H}
\end{equation}
where $\sigma_i^{\alpha}$ is the Pauli matrix at site $i$. Although we
treat spin-$1/2$ systems in this paper, one can straightforwardly
generalize all the results to spin-$S$ systems. There is no restriction
in the spatial dimensionality or lattice structure.  The exchange
interaction $J_{ij}$ is a quenched random variable. One of the useful
probability distributions for $J_{ij}$ is the binary distribution
\begin{equation}
 P(J_{ij}) = p\, \delta(J_{ij}-J) +(1-p)\, \delta(J_{ij}+J) .
\end{equation}
It is convenient for later arguments to rewrite this distribution as
\begin{equation}
 P(J_{ij}) = \frac{{\rm e}^{K_p \tau_{ij}}}{2\cosh K_p}, \qquad 
 K_p= \frac{1}{2}\log \frac{p}{1-p} ,
 \label{P_pm}
\end{equation}
where $\tau_{ij} = J_{ij}/J$ is the sign of the exchange interaction
$J_{ij}$. Another useful distribution is the Gaussian
distribution
\begin{equation}
 P(J_{ij}) = \frac{1}{\sqrt{2\pi J^2}}
  \exp\left(-\frac{(J_{ij}-J_0)^2}{2J^2}\right),
  \label{P_gauss}
\end{equation}
where $J_0$ and $J^2$ denote the average and variance, respectively.

For quantum spin systems, the classical gauge transformation, which
simultaneously changes the sign of all components, is not valid because
the commutation rule $[\sigma_i^x,\sigma_i^y]=2{\rm i}\sigma_i^z$ is
changed to $[\sigma_i^x,\sigma_i^y]=-2{\rm i}\sigma_i^z$. Thus we define a
gauge transformation for spins using a unitary operator as
\begin{equation}
 U : \sigma_i^\alpha \rightarrow 
  G \sigma_i^\alpha G^{-1} , \qquad
  G = \prod_i G_i, \qquad
  G_i = 
  \begin{cases}
   1_i & (\xi_i = +1) \\
   \exp\left(-\displaystyle\frac{{\rm i}\pi}{2} \sigma_i^x \right) 
   & (\xi_i=-1)
  \end{cases} ,
  \label{TIM_G}
\end{equation}
where $\xi_i$ is a classical gauge variable at site $i$ and takes two
values $\pm 1$.  If $\xi_i=-1$, $\sigma_i^{y,z} \rightarrow
-\sigma_i^{y,z}$ and $\sigma_i^x \rightarrow \sigma_i^x$. Equivalently
we can write
\begin{equation}
 U : (\sigma_i^x, \sigma_i^y, \sigma_i^z) \rightarrow
  (\sigma_i^x, \xi_i \sigma_i^y, \xi_i \sigma_i^z) .
\end{equation}
A difference of gauge transformations between quantum and
classical systems is the transformation rule of $\sigma_i^x$.

The gauge transformation for the bond variables $\{J_{ij}\}$ is the same
as in classical systems, namely
\begin{equation}
 V : J_{ij} \rightarrow J_{ij} \xi_i \xi_j. \label{V}
\end{equation}
The transverse-field term in eq. (\ref{TIM_H}) does not change by the
gauge transformation.

The Hamiltonian (\ref{TIM_H}) is clearly invariant under the successive
operations of $V$ and $U$: $(UV)H=H$.  However, the distribution
function of bond configuration is changed, for the $\pm J$ Ising model,
as
\begin{equation}
  P(J_{ij}) \rightarrow 
   \frac{{\rm e}^{K_p \tau_{ij}\xi_i\xi_j}}{2\cosh K_p}.
   \label{P_pm2}
\end{equation}
It is important for the following argument that this transformed
distribution is proportional to the Boltzmann factor of a classical
system.  Similarly, the Gaussian distribution (\ref{P_gauss}) is changed as
\begin{equation}
 P(J_{ij}) \rightarrow \frac{1}{\sqrt{2\pi J^2}}
  \exp\left({-\frac{J_{ij}^2+J_0^2}{2J^2}}\right)
  \exp\left({\frac{J_0}{J^2}J_{ij}\xi_i\xi_j}\right) .
\end{equation}
To simplify the arguments, we focus on the binary distribution
(\ref{P_pm}) and (\ref{P_pm2}), hereafter. It is straightforward to
apply the same methods to the Gaussian distribution.

\subsection{Quantum gauge glass}
Next, we consider the quantum gauge glass (QGG). Similarly to the
transverse Ising model, the gauge transformation is defined by the
rotation operator.

To properly define the quantum version of gauge glass, let us first
consider the Hamiltonian of the classical gauge glass (CGG),
\begin{equation}
 H_{\rm cl} = -J\sum_{\langle ij \rangle}\cos (\phi_i-\phi_j-\omega_{ij}) .
\end{equation}
This Hamiltonian can be rewritten using the spin vector composed of $x$
and $y$ components
$\mib{S}_i
=\left(\begin{smallmatrix}
	S_i^x\\
	S_i^y
       \end{smallmatrix} \right)$
and rotational matrix in the $XY$ plane,
$R(\theta)=\left(\begin{smallmatrix}
		  \cos\theta & -\sin\theta \\
		  \sin\theta & \cos\theta
		 \end{smallmatrix}\right)$
as
\begin{equation}
 H_{\rm cl} = -J \sum_{\langle ij \rangle} {^t\mib{S}_i} R(\omega_{ij})
  \mib{S}_j.\label{Hc_gg} 
\end{equation}
Thus this model can be quantized straightforwardly by replacing the
elements of the above spin vectors by the Pauli matrices. The
Hamiltonian of the QGG is therefore written explicitly as
\begin{equation}
 H=-J\sum_{\langle ij \rangle}\left\{
 \cos\omega_{ij}\left(\sigma_i^x \sigma_j^x
		 +\sigma_i^y \sigma_j^y \right)
 -\sin\omega_{ij}\left(\sigma_i^x \sigma_j^y
		  -\sigma_i^y \sigma_j^x \right)\right\} .
 \label{H_qgg}
\end{equation}
The phase factor $\omega_{ij}\in [0,2\pi)$ is a quenched random variable
whose probability distribution is of cosine type
\begin{equation}
 P(\omega_{ij})=\frac{{\rm e}^{K_p\cos\omega_{ij}}}{2\pi I_0(K_p)} ,
  \label{P_omega}
\end{equation}
a periodic Gaussian (Villain) type
\begin{equation}
 P(\omega_{ij})=\sqrt{\frac{K_p}{2\pi}}\sum_{n=-\infty}^{\infty}
  \exp\left(-\frac{K_p(\omega_{ij}-2\pi n)^2}{2}\right)
\end{equation}
or a binary type
\begin{equation}
 P(\omega_{ij})= p\,\delta(\omega_{ij})+(1-p)\delta(\omega_{ij}-\pi).
\end{equation}

Equation (\ref{H_qgg}) is a special case of the $XY$ model with
Dzyaloshinskii-Moriya interactions. This Hamiltonian is written as
\begin{equation}
 H = -J\sum_{\langle{ij}\rangle}\mib{\sigma}_i \cdot \mib{\sigma}_j 
  -\sum_{\langle{ij}\rangle} J_{ij} 
  (\mib{\sigma}_i\times\mib{\sigma}_j)_z,
\end{equation}
where the second term is the random Dzyaloshinskii-Moriya interaction.
If we set new parameters,
\begin{equation}
 \tilde{J}_{ij} = \sqrt{J^2+J_{ij}^2} , \quad
  \omega_{ij} = -\tan^{-1} \left(\frac{J_{ij}}{J}\right) ,
\end{equation} 
the above Hamiltonian is rewritten as
\begin{equation}
  H=-\sum_{\langle ij \rangle} \tilde{J}_{ij }\left\{
 \cos\omega_{ij}\left(\sigma_i^x \sigma_j^x
		 +\sigma_i^y \sigma_j^y \right)
 -\sin\omega_{ij}\left(\sigma_i^x \sigma_j^y
		  -\sigma_i^y \sigma_j^x \right)\right\}.
\end{equation}
This is equal to eq. (\ref{H_qgg}) except that the interaction
depends on bonds. 

In the CGG, the gauge transformation for spins is defined by the shift
of spin variables as $\phi_i\rightarrow\phi_i-\psi_i$, where $\psi_i\in
[0,2\pi)$ is the gauge variable. Using the same notation as in
eq. (\ref{Hc_gg}), this transformation is expressed as
\begin{equation}
 U : \mib{S}_i \rightarrow R(-\psi_i)\mib{S}_i .
\end{equation}
Thus we use this definition of gauge transformation for the QGG. Using a
rotational operator on the Hilbert space, we define
\begin{equation}
 U : \mib{\sigma}_i \rightarrow G \mib{\sigma}_i G^{-1} \quad
 G = \prod_i \exp\left(-\frac{{\rm i} \psi_i}{2}\sigma_i^z \right).
\end{equation}
The transformation rule for the transposed vector $^t \mib{\sigma}_i$ is
defined as
\begin{equation}
 U : {^t\mib{\sigma}_i} \rightarrow {^t \mib{\sigma}_i} R(\psi_i) =
  G\, {^t\!\mib{\sigma}_i} G^{-1} .
\end{equation}
The gauge transformation of random variables is the same as in the
classical case,
\begin{equation}
 V : \omega_{ij} \rightarrow \omega_{ij} -\psi_i +\psi_j.
\end{equation}
Under the gauge transformation $UV$, the Hamiltonian is invariant because
\begin{equation}
 (UV) H = -J \sum_{\langle ij \rangle} {^t\mib{\sigma}_i} R(\psi_i) 
  R(\omega_{ij}-\psi_i+\psi_j) R(-\psi_j) \mib{\sigma}_j = H,
\label{invQGG}
\end{equation}
where we used the property of rotation matrices,
$R(\psi)R(\phi)=R(\psi+\phi)$. The probability distribution (\ref{P_omega})
is changed as
\begin{equation}
 P(\omega_{ij}) \rightarrow 
  \frac{{\rm e}^{K_p\cos(\omega_{ij}-\psi_i+\psi_j)}}{2\pi I_0(K_p)}.
\end{equation}
This transformed distribution is proportional to the Boltzmann factor
for the CGG. If we choose the Gaussian or binary type, the Boltzmann
factor for the Villain or $\pm J$ model appears, respectively.

\section{Identity for Gauge Invariant Operators}
The gauge symmetry of the Hamiltonian yields a useful identity for gauge
invariant operators. First, let us suppose that the system was initially
in the perfect FM state $|F_z\rangle$ in the transverse Ising
model. This state appears in the FM limit, $T=0$, $p=1$,
$h=0$. The gauge transformation operator $G$ defined in eq. (\ref{TIM_G})
changes this state as
\begin{equation}
 G |F_z\rangle = |\xi\rangle, \qquad
  |\xi\rangle = |\xi\rangle_1 |\xi\rangle_2 
  \cdots |\xi\rangle_N.
  \label{Fz}
\end{equation}
If $\xi_i=+1$, $|\xi\rangle_i$ denotes the state with up spin in
the $z$ direction, and if $\xi_i=-1$, the spin at site $i$ is down.

Using the property of $|F_z\rangle$ in (\ref{Fz}), we prove the
following identity for a gauge-invariant operator $Q$ which satisfies
$Q=(UV)Q$ or equivalently $VQ=G^{-1} Q G$,
\begin{equation}
 \left[\left\langle Q \right\rangle_{F_z}\right]
  =\left[\left\langle Q \right\rangle_{\rho_{\rm cl}(K_p)}\right], \label{Q}
\end{equation}
where $\left\langle \cdot \right\rangle_{\rho_{\rm cl}(K_p)}$ is the expectation value
for the classical equilibrium state on the Nishimori line (NL), that is,
\begin{equation}
 \left\langle Q \right\rangle_{\rho_{\rm cl}(K_p)}
  ={\rm Tr}\, \rho_{\rm cl}(K_p) Q, \qquad
  \rho_{\rm cl}(K_p)
  =\frac{{\rm e}^{K_p\sum_{{ij}}\tau_{ij}\sigma_i^z \sigma_j^z}}
  {{\rm Tr}\, {\rm e}^{K_p\sum_{{ij}}\tau_{ij}\sigma_i^z \sigma_j^z}}.
\end{equation}

To prove the identity (\ref{Q}), we apply the gauge transformation for
the configuration of randomness of eq. (\ref{V}) appearing on left-hand
side of eq. (\ref{Q}). This operation does not change its value because
the transformation $V$ of eq. (\ref{V}) only changes the order of the
summation over $\tau_{ij}$. Thus the left-hand side of eq. (\ref{Q}) is
rewritten as
\begin{equation}
 \left[\left\langle Q\right\rangle_{F_z}\right]
 = \sum_{\{\tau_{ij}\}}
  \frac{{\rm e}^{K_p\sum\tau_{ij}\xi_i \xi_j}}
  {(2\cosh K_p)^{N_B}} \langle F_z | VQ | F_z \rangle
 = \sum_{\{\tau_{ij}\}} 
  \frac{{\rm e}^{K_p\sum\tau_{ij}\xi_i \xi_j}}
  {(2\cosh K_p)^{N_B}} \langle \xi | Q | \xi \rangle,
\end{equation}
where we used the assumption that the operator $Q$ is gauge invariant,
$VQ = G^{-1}QG$.
Since the expectation value on the left-hand side does not depend on
$\xi$, the summation over $\xi$ and division by $2^N$ yield
\begin{equation}
 \left[\left\langle Q \right\rangle_{F_z}\right]
 = \sum_{\{\tau_{ij}\}} 
  \frac{1}{2^N (2\cosh K_p)^{N_B}} \sum_{\{\xi\}}
  {\rm e}^{K_p\sum\tau_{ij}\xi_i \xi_j} 
  \langle \xi | Q |\xi\rangle.
\end{equation}
The last part of the right-hand side is rewritten
in terms of $\rho_{\rm cl}(K_p)$ as
\begin{equation}
  \sum_{\{\xi\}} {\rm e}^{K_p\sum\tau_{ij}\xi_i \xi_j}
  \langle\xi| Q | \xi\rangle
  =\left(\sum_{\{\xi\}} {\rm e}^{K_p\sum\tau_{ij}\xi_i \xi_j}\right) 
  \left({\rm Tr} \rho_{\rm cl}(K_p) Q\right).
\end{equation}
Therefore, we obtain
\begin{equation}
  \left[\left\langle Q\right\rangle_{F_z}\right]
  = \sum_{\{\tau_{ij}\}} 
  \frac{\sum_{\xi} {\rm e}^{K_p\sum\tau_{ij}\xi_i \xi_j}}
  {2^N (2\cosh K_p)^{N_B}} {\rm Tr} \rho_{\rm cl}(K_p) Q.
\end{equation}
Since ${\rm Tr}\rho_{\rm cl}(K_p) Q$ is invariant under the transformation $V$,
this is identical to the right-hand side of eq. (\ref{Q})

The above result can be generalized to the case that the transverse
field $h(t)$ depends on time, following the classical example. \cite{O1}
We consider the zero-temperature time evolution following the
Schr\"{o}dinger equation. Using the time ordered product, the time
evolution operator is written as
\begin{equation}
 U_t = T \exp\left(-{\rm i}\int_{0}^{t} H(t') {\rm d}t' \right) .
\end{equation}
Since the time dependence of the Hamiltonian does not invalidate the gauge
symmetry, this operator is also gauge invariant
\begin{equation}
 (UV)U_t = G(V U_t) G^{-1} = U_t .
\end{equation}

Examples of the gauge invariant operator include the transverse
magnetization $\sigma_i^x(t)= U_t^{\dagger} \sigma_i^x U_t$, the
autocorrelation function $\sigma_i^z(0) \sigma_i^z(t)$ and the
interaction term $H_0(t)$ of the Hamiltonian (the first term on the
right-hand side of eq. (\ref{TIM_H})).

For the QGG, we can prove a similar identity
\begin{equation}
 \left[\left\langle Q\right\rangle_{F_x}\right]
  = \left[\left\langle Q\right\rangle_{\rho_{\rm cl}(K_p)}\right].
  \label{Q_qgg}
\end{equation}
Here one should note that $\rho_{\rm cl}(K_p)$ is different from the normal
density operator. If we choose the cosine-type distribution (\ref{P_omega}),
$\rho_{\rm cl}(K_p)$ is defined by the Boltzmann factor for the CGG as
\begin{equation}
 \rho_{\rm cl}(K_p) = \frac{{\rm Tr}_{\psi} {\rm e}^{K_p \sum_{ij} 
  \cos(\omega_{ij}-\psi_i+\psi_j)} 
  |{\psi}\rangle \langle{\psi}| }
  {{\rm Tr}_{\psi} {\rm e}^{K_p \sum_{ij} 
  \cos(\omega_{ij}-\psi_i+\psi_j)}} ,
\end{equation}
where $|\psi\rangle = G |F_x\rangle$ and $\text{Tr}_{\psi}$ stands for
integration over $\psi_i$ from $0$ to $2\pi$. Since the state vector $|
\psi\rangle$ does not diagonalize the Hamiltonian for the quantum gauge
glass, $\rho_{\rm cl}(K_p)$ is not equal to the density operator
\begin{equation}
 \rho(K_p) = \frac{{\rm e}^{-\beta H}}{{\rm Tr}\, {\rm e}^{-\beta H}} .
\end{equation}

The identities (\ref{Q}) and (\ref{Q_qgg}) show that the expectation
value of gauge invariant operator in the FM limit is equal to
the one in the classical equilibrium state on the NL. The equivalence of
the two states has already been proved in the dynamical gauge theory
for classical systems \cite{O1,O2}.  The present results are
generalization of these dynamical cases to quantum systems. Note that
the zero-temperature time evolution for quantum systems is deterministic
in contrast to the stochastic dynamics for the classical SG.

\section{Correlation Function and Order Parameter}
Using the identities proved in the previous section, we can derive a
class of inequalities for the correlation function. First, we treat
the transverse Ising model. Since the correlation function is not
invariant under the gauge transformation, let us consider the following
gauge-invariant quantity
\begin{equation}
 Q= \sigma_i^z \sigma_j^z \langle \sigma_i^z \sigma_j^z \rangle_{K,h},
\end{equation}
where $\langle\cdot\rangle_{K,h}$ denotes thermal average with temperature
$\beta^{-1}=J/K$ under a transverse field $h$. Substitution of the above
quantity into eq. (\ref{Q}) yields
\begin{equation}
 \left[\left\langle \sigma_i^z \sigma_j^z\right\rangle_{K,h}\right]
  =\left[\left\langle \xi_i \xi_j\right\rangle^{\rm cl}_{K_p}
    \left\langle \sigma_i^z \sigma_j^z\right\rangle_{K,h}\right].
\end{equation}
Here $\langle \xi_i \xi_j \rangle_{K_p}^{\rm cl}$ is the correlation function
for the classical Ising system with the same configuration
$\{\tau_{ij}\}$ and no external field. By taking the absolute value of
both sides of this equation, we find
\begin{equation}
 \left|\left[\left\langle \sigma_i^z \sigma_j^z
	      \right\rangle_{K,h}\right]\right|
  \leq\left[\left|\left\langle \xi_i \xi_j
		  \right\rangle^{\rm cl}_{K_p}\right|\right],
  \label{ineq1}
\end{equation}
where we used the fact that correlation function $\left\langle{\sigma_i^z
\sigma_j^z}\right\rangle_{K,h}$ does not exceed unity. Similarly, we can
prove
\begin{equation}
 \left|\left[{\rm sgn}\left(\left\langle \sigma_i^z \sigma_j^z
      \right\rangle_{K,h} \right)\right]\right|
  \leq \left[\left|\langle
	      \xi_i \xi_j \rangle_{K_p}^{\rm cl} \right| \right]
\end{equation}
\begin{equation}
 \left[\frac{ \langle \xi_i \xi_j \rangle_{K_p}^{\rm cl}}
  {\left\langle \sigma_i^z \sigma_j^z \right\rangle_{K,h}}\right]
 = \left[\frac{1}
  {\left\langle \sigma_i^z \sigma_j^z \right\rangle_{K,h}}\right]
 \geq 1.
 \label{ineq2}
\end{equation}

\begin{figure}
 \begin{center}
  \includegraphics{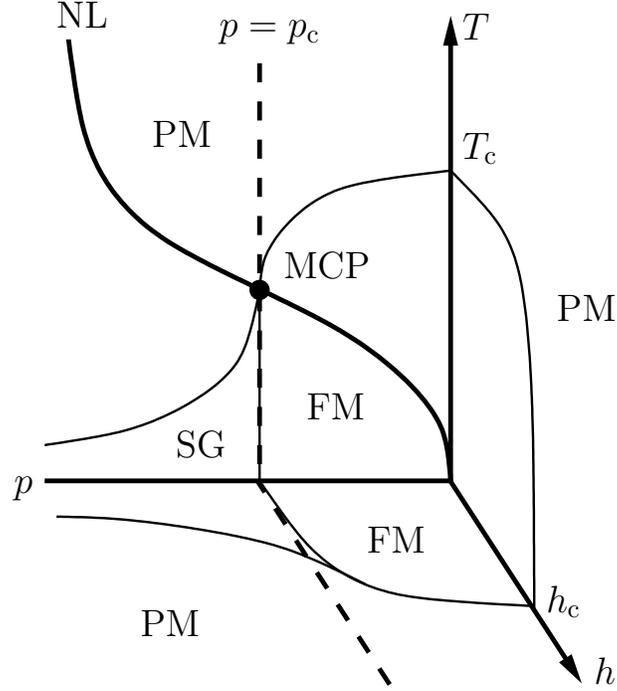} 
  \caption{The phase diagram of the transverse
  Ising model. The paramagnetic (PM), the ferromagnetic (FM) and spin
  glass (SG) phases meet at the multicritical point (MCP) in the plane
  $h=0$. The dashed line at $p=p_{\rm c}$ sets a bound for the existence of the
  FM phase also for $h\neq 0$.}  \label{TIM}
 \end{center}
\end{figure}

If the probability $p$ of the FM interaction is less than the
critical probability $p_{\rm c}$ at the multicritical point for the classical
Ising system, the right-hand side of the inequality (\ref{ineq1})
vanishes in the limit $\left| i-j \right|\rightarrow\infty$. Thus the
correlation function for the transverse Ising model on the left-hand
side is also equal to zero. Therefore the region of the FM
phase is restricted to the range $p > p_{\rm c}$ (Fig. \ref{TIM}). Since the
transverse field represents quantum fluctuations, the correlation
function for the transverse Ising model should be reduced from classical
system with $h=0$, which is the physical origin of the above-mentioned
restriction.

Next, let us consider how to define a correlation function of the QGG
which has convenient properties for the gauge theory. In the CGG, a
useful correlation function is defined in terms of an exponential
function as ${\rm e}^{{\rm i}(\phi_i-\phi_j)}$. This is rewritten using
the notation of eq. (\ref{H_qgg}) as
\begin{equation}
 {\rm e}^{{\rm i}(\phi_i-\phi_j)} = \mib{S}_i\cdot\mib{S}_j 
 -{\rm i}\,(\mib{S}_i \times \mib{S}_j)_z
 = {^t\mib{S}_i}\mib{S}_j -{\rm i}\, {^t\mib{S}_i} R(-\pi/2) \mib{S}_j .
\label{CF_gg}
\end{equation}
This motivates us to define a correlation operator $\gamma_{ij}$ for the
quantum gauge glass as,
\begin{equation}
 \gamma_{ij}= (\sigma_i^x \sigma_j^x+\sigma_i^y \sigma_j^y) 
  -{\rm i} (\sigma_i^x \sigma_j^y-\sigma_i^y \sigma_j^x).\label{CF_qgg}
\end{equation}
The gauge transformation $U$ changes this operator according to
\begin{equation}
 U \gamma_{ij} = G \gamma_{ij} G^{-1}
 = {\rm e}^{-{\rm i}(\psi_i-\psi_j)} \gamma_{ij} .
 \label{UCF}
\end{equation}
The first factor is the same as the correlation function appearing in
the classical gauge glass except for a minus sign.

Usually, the correlation function for $XY$-like systems is defined as
the expectation value of $\sigma_i^x \sigma_j^x+\sigma_i^y \sigma_j^y$,
which corresponds to $\cos(\phi_i-\phi_j)$ for classical
systems. However, this expression is not useful for the gauge theory
because it does not separate into gauge variables and spin operators
after transformation by $U$ as
\begin{equation}
\begin{split}
 U(\sigma_i^x \sigma_j^x+\sigma_i^y \sigma_j^y)
 = \cos(\psi_i-\psi_j)(\sigma_i^x \sigma_j^x+\sigma_i^y \sigma_j^y)
 -\sin(\psi_i-\psi_j)(\sigma_i^x \sigma_j^y-\sigma_i^y \sigma_j^x).
\end{split}
\end{equation}
In addition, one can easily prove that 
\begin{equation}
 \left[\left\langle \gamma_{ij}\right\rangle_K \right]=
  \left[\left\langle \sigma_i^x \sigma_j^x
	 +\sigma_i^y \sigma_j^y \right\rangle_K \right]
\end{equation}
since the QGG is invariant under the following transformation:
\begin{equation}
 (\sigma_i^x, \sigma_i^y, \sigma_i^z) \rightarrow
 (\sigma_i^y, -\sigma_i^x, \sigma_i^z), \qquad
 \omega_{ij} \rightarrow -\omega_{ij} .
\end{equation}
Thus we choose to discuss the correlation operator (\ref{CF_qgg}).

To derive an inequality similar to eq. (\ref{ineq1}), it is useful to
consider
\begin{equation}
 Q = \gamma_{ij}^\dagger \langle\gamma_{ij}\rangle_K .
\end{equation}
It is easy to prove that this quantity is gauge invariant because of the
property of $\gamma_{ij}$ given in eq. (\ref{UCF}). Substituting
this quantity into eq. (\ref{Q_qgg}), we obtain
\begin{equation}
 \left[\left\langle \gamma_{ij}\right\rangle_K \right]
  =\left[ \left\langle {\rm e}^{-{\rm i}(\psi_i-\psi_j)}
	  \right\rangle_{K_p}^{\rm cl} 
    \left\langle \gamma_{ij}\right\rangle_K \right],
\label{gmgmQGG}
\end{equation}
where $\langle \cdot \rangle^{\rm cl}_{K_p}$ stands for the expectation value
for the CGG on the NL.  By taking the absolute value of both sides of
this equation, we find
\begin{equation}
 \left|\left[\left\langle\gamma_{ij}
	     \right\rangle_K \right]\right|
  \leq
  \left[\left|\left\langle {\rm e}^{-{\rm i}(\psi_i-\psi_j)}
	       \right\rangle^{\rm cl}_{K_p}\right|\right].
  \label{QGG_ineq}
\end{equation}
Similarly, we obtain
\begin{equation}
 \left|\left[\text{sgn}(\left\langle \gamma_{ij}
			\right\rangle_K)\right]\right|
 \leq \left[\left|\left\langle {\rm e}^{-{\rm i}(\psi_i-\psi_j)}
		   \right\rangle^{\rm cl}_{K_p}\right|\right],
\end{equation}
\begin{equation}
 \left[\frac{\left\langle {\rm e}^{{\rm i}(\psi_i-\psi_j)}
	     \right\rangle^{\rm cl}_{K_p}}
  {\left\langle \gamma_{ij}\right\rangle_K}\right]
 =\left[\frac{1}{\left\langle \gamma_{ij}\right\rangle_K}\right]
 \geq 1.
\end{equation}

Since the lower critical dimension $d_l$ is two for continuous spin
systems, for $d>2$, we expect a FM phase to exist at low
temperature under small randomness. In this case, a two-spin correlation
function tends to the square of magnetization when the two spins are
sufficiently separated,
\begin{equation}
 \left[\left\langle\gamma_{ij}\right\rangle_K \right] 
  \rightarrow m(K,K_p)^2,
  \qquad \left|i-j\right| \rightarrow \infty.
\end{equation}
The right-hand side of inequality (\ref{QGG_ineq}) is estimated as
follows:
\begin{equation}
 \begin{split}
  \left[\left|\left\langle {\rm e}^{-{\rm i}(\psi_i-\psi_j)}
  \right\rangle^{\rm cl}_{K_p}\right|\right]^2
  &\leq
  \left[\left|\left\langle {\rm e}^{-{\rm i}(\psi_i-\psi_j)}
  \right\rangle^{\rm cl}_{K_p}\right|^2\right] \\
  &\rightarrow
  \left[\left|\left\langle {\rm e}^{-{\rm i}\psi_i}
  \right\rangle^{\rm cl}_{K_p}\right|^2\right]
  \left[\left|\left\langle {\rm e}^{{\rm i}\psi_j}
  \right\rangle^{\rm cl}_{K_p}\right|^2\right]
  \qquad \left(\left|i-j\right|\rightarrow\infty\right) \\
  &= q^{\rm cl}(K_p,K_p)^2 = m^{\rm cl}(K_p,K_p)^2,
 \end{split}
\end{equation}
where we used the identity $m^{\rm cl}=q^{\rm cl}$ resulting from the gauge theory on the
NL for the CGG. Therefore we obtain
\begin{equation}
 m(K,K_p)^2 \leq m^{\rm cl}(K_p,K_p).
\label{mmQGG}
\end{equation}
If the parameter $K_p$ is smaller than the critical point $K_{p_{\rm c}}^{\rm cl}$
for the classical system, the right-hand side vanishes. Consequently,
the FM phase for the quantum gauge glass lies in the region
satisfying $K_p > K_{p_{\rm c}}^{\rm cl}$. This result is consistent with the
intuitive picture that quantum effects reduce long-range order.

If the spatial dimensionality of the system is equal to the lower
critical dimension, $d=2$, there is no long-range order but quasi
long-range order. The Kosterlitz-Thouless (KT) phase \cite{KT} exists
when both $K$ and $K_p$ are sufficiently large. The ordering tendency of
the KT phase is observed by the correlation length $\xi$ in the
paramagnetic (PM) phase as
\begin{gather}
 \left[\left\langle\gamma_{ij}\right\rangle_K \right]
 \sim {\rm e}^{-|i-j|/\xi_m(K,K_p)} ,\\
 \left[\left|\left\langle {\rm e}^{-{\rm i}(\psi_i-\psi_j)}
 \right\rangle^{\rm cl}_{K_p}\right|^2 \right]
 = \left[ \left\langle {\rm e}^{-{\rm i}(\psi_i-\psi_j)}
 \right\rangle^{\rm cl}_{K_p} \right]
 \sim {\rm e}^{-|i-j|/\xi^{\rm cl}_m(K_p,K_p)} .
\end{gather}
From the square of the inequality (\ref{QGG_ineq}), the limit 
$\left|i-j\right|\rightarrow\infty$ yields
\begin{equation}
 \xi_m(K,K_p) \leq 2 \xi^{\rm cl}_m(K_p,K_p).
\label{xxQGG}
\end{equation}
Thus, the KT phase for the QGG is also restricted to the
region $K_p>K_{p_{\rm c}}^{\rm cl}$.

\section{Ground State Property of Quantum Gauge Glass}
Next we consider the ground state property of the QGG.  It is necessary
to consider the transformation rule for eigenstates of the Hamiltonian.
Let us denote an eigenstate of the Hamiltonian with the eigenvalue $x$
by $|x;{\mib \omega}\rangle$,
\begin{equation}
H |x;{\mib \omega}\rangle = x~|x;{\mib \omega}\rangle. 
\end{equation}
Since the invariance of the Hamiltonian (\ref{invQGG}) 
can be rewritten as
\begin{equation}
G(VH)G^{-1} =H,
\end{equation}
we have an eigen-value equation
\begin{equation}
(VH)~G^{-1}|x;{\mib \omega}\rangle = x~G^{-1}|x;{\mib \omega}\rangle.
\end{equation}
Note that the effect of the operator $V$ is restricted to the inside of the
brackets $(\cdots)$.
The state $G^{-1}|x;{\mib \omega}\rangle$ is an eigenstate of 
the gauge-transformed Hamiltonian $(VH)$. 
Therefore, one can derive the transformation rule of the eigenstate
\begin{equation}
V |x;{\mib \omega}\rangle = G^{-1}|x;{\mib \omega}\rangle. 
\end{equation}
This rule is derived straightforwardly, if the state is not degenerate.
When the state is degenerate, the rule is not unique.
However, it is reasonable and has no problem, if we use this rule as the
transformation rule.

Now, we consider the ground sate.
Let us denote the ground state of the
Hamiltonian by $|{\rm g};{\mib \omega}\rangle$.
The above transformation rule leads to the gauge transformation of
the average of any operator $Q$ in the ground state;
\begin{equation} 
V \langle {\rm g};{\mib \omega}|Q|{\rm g};{\mib \omega}\rangle
=\langle {\rm g};{\mib \omega}| G(VQ)G^{-1}|{\rm g};{\mib \omega}\rangle.
\end{equation}
The thermal average $\langle\cdots\rangle_K$ in equations derived 
in the previous section can be replaced by the ground state expectation
value.
For example,
\begin{equation}
 \left[
\left\langle {\rm g};{\mib \omega}|\gamma_{ij}
|{\rm g};{\mib \omega}\right\rangle
\right]
  =\left[ \left\langle {\rm e}^{-{\rm i}(\psi_i-\psi_j)}
	  \right\rangle_{K_p}^{\rm cl} 
\left\langle {\rm g};{\mib \omega}|\gamma_{ij}|{\rm g};{\mib \omega}\right\rangle
\right]
\end{equation}
is derived instead of eq.\ (\ref{gmgmQGG}),
which provides the inequality for the order parameters in the ground state,
\begin{equation}
m(\infty,K_p)^2 \le m^{\rm cl}(K_p, K_p)
\label{mmGQGG}
\end{equation} 
instead of eq.\ (\ref{mmQGG}), and 
\begin{equation}
\xi_m(\infty,K_p) \le 2\xi_m^{\rm cl}(K_p, K_p)
\label{xxGQGG}
\end{equation} 
instead of eq.\ (\ref{xxQGG}).

In two dimensions, it has been shown that the FM long range order
exists in the ground state of the pure quantum $XY$ model 
($K_p =+\infty$).\cite{KLS}
However, in the disordered regime ($K_p<+\infty$),
the FM order must disappear since the FM order in the CGG model,
the right-hand side of the inequality (\ref{mmGQGG}), does not exist.
The only possibility in this regime is the existence of the KT phase,
which is consistent with the inequality (\ref{xxGQGG}).

\section{Phase Diagram of Quantum Mattis Model}
In this section, we introduce and discuss the properties of the quantum
$XY$ Mattis model which has no frustration. Using the gauge
transformation, one can obtain the phase diagram for this model
explicitly. One of the phase boundaries is determined by the critical
point of the pure quantum system and the other by that of the classical
one. This is an important difference from the classical non-frustrated
systems.\cite{O3}

Let us locate a quenched random variable $\omega_i$ at each site and
define the phase factor $\omega_{ij}$ as
$\omega_{ij}=\omega_j-\omega_i$. The Hamiltonian for the quantum Mattis
model is defined in terms of the pure quantum $XY$ model, $H_0$, that is,
\begin{equation}
 H = G_{\mib \omega}H_0 G_{\mib \omega}^{-1}, \qquad
  G_{\mib \omega}=\prod_i 
  \exp\left(-\frac{{\rm i}\omega_i}{2} \sigma_i^z\right) .
  \label{H_m}
\end{equation}
Thus, the ground-state energy is always the same as that of $H_0$ and
the ground state is obtained by operating $G_{\mib \omega}$ on the
ground state of the pure system. This system has no frustration in this
sense. It is easy to show that the Hamiltonian for this model is invariant
under the gauge transformation, $(UV)H=H$, where the gauge transformation
for the configuration is defined as $\omega_i\rightarrow\omega_i-\psi_i$.

Using eqs. (\ref{UCF}) and (\ref{H_m}), we immediately obtain
\begin{equation}
 \left\langle \gamma_{ij}\right\rangle_K
  = {\rm e}^{{\rm i}(\omega_i-\omega_j)}
  \left\langle\gamma_{ij}\right\rangle_{0,K},
  \label{M_cf}
\end{equation}
where the angular brackets on the right-hand side denote the thermal
average with respect to $H_0$. Note that the correlation function for
the pure system $\left\langle\gamma_{ij}\right\rangle_{0,K}$ does not
depend the quenched variable $\omega_i$.  Here, we assume that the
distribution function for the quenched random variable is proportional
to the Boltzmann factor of the pure classical $XY$ model,
\begin{equation}
 P({\mib \omega})=
  \frac{{\rm e}^{K_p \sum_{ij}\cos(\omega_i-\omega_j)}}{Z_0^{\rm cl}(K_p)}.
\end{equation}
Hereafter, the subscript $0$ and the superscript cl stand for pure and
classical systems, respectively. Then the configuration average is equal
to the thermal average for the pure classical $XY$ model with coupling
$K_p$. Consequently, we find
\begin{equation}
 \left[ \left\langle \gamma_{ij}\right\rangle_K\right]
  =\left\langle {\rm e}^{{\rm i}(\omega_i-\omega_j)}\right\rangle_{0,K_p}^{\rm cl}
  \left\langle\gamma_{ij}\right\rangle_{0,K}.
  \label{MaCF}
\end{equation}
Similarly, from eq. (\ref{M_cf}), the
spin-glass correlation function satisfies
\begin{equation}
 \left[\left|\left\langle \gamma_{ij}\right\rangle_K \right|^2\right]
  =\left|\left\langle\gamma_{ij}\right\rangle_{0,K}\right|^2.
  \label{MaCF2}
\end{equation}

\begin{figure}
 \begin{center}
  \includegraphics{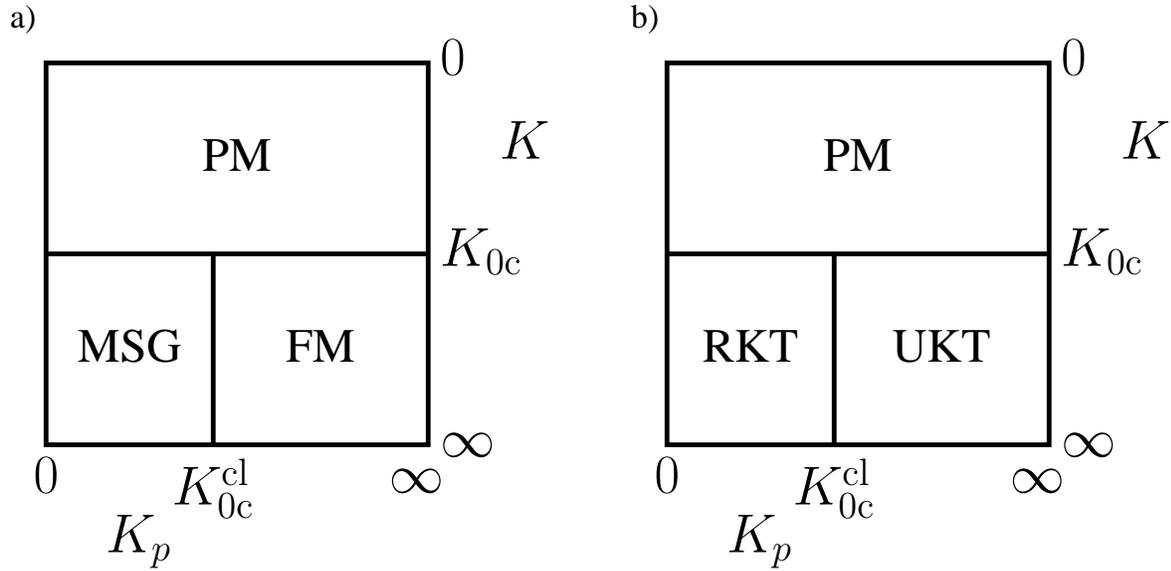} \caption{(a) The phase diagram for the
  quantum $XY$ Mattis model in $d>2$. There are three kinds of phases,
  the PM, the FM and the SG. (b) The phase diagram in $d=2$. There exist
  three kinds of phases, the PM, the uniform KT (UKT) and the random KT
  (RKT).}  \label{Mpd}
 \end{center}
\end{figure}

Taking the limit $\left|i-j\right|\rightarrow\infty$ of eqs. (\ref{MaCF})
and (\ref{MaCF2}), we obtain
\begin{equation}
 m(K,K_p)=m_0^{\rm cl}(K_p) m_0(K),
\end{equation}
\begin{equation}
 q(K,K_p)=m_0(K)^2.
\end{equation}
For $d>2$, there are three phases, PM, FM
and Mattis spin-glass (MSG) phases. Figure \ref{Mpd}(a) shows the phase
diagram. The phase boundary between PM and other phases is at $K=K_{0\rm c}$
and the one between FM and MSG is at $K=K_{0\rm c}^{\rm cl}$.

If $d=2$, the correlation length determines the phase structure. From
eqs. (\ref{MaCF}) and (\ref{MaCF2}), we find
\begin{equation}
 \frac{1}{\xi_m(K,K_p)}=\frac{1}{\xi_0^{\rm cl}(K_p)}+\frac{1}{\xi_0(K)}
\end{equation}
\begin{equation}
 \xi_q(K,K_p)=\frac{\xi_0(K)}{2},
\end{equation}
where $\xi_q(K,K_p)$ denotes the spin-glass correlation length.
Thus, similarly to the $d>2$ case, three phases exist: 
(i) $\xi_m<\infty$ and $\xi_q<\infty$: paramagnetic phase (PM),
(ii) $\xi_m=\infty$ and $\xi_q=\infty$: uniform KT phase (UKT) and
(iii) $\xi_m<\infty$ and $\xi_q=\infty$: random KT phase (RKT).
The phase diagram is shown in Fig. \ref{Mpd}(b).

We note again that the location of the horizontal phase boundary is
determined by the critical point of the quantum pure system, $K_{0\rm c}$,
and the vertical one comes from that of the classical pure system,
$K_{0\rm c}^{\rm cl}$.

\section{Conclusions}
In this paper, we have investigated quantum spin glasses using the gauge
theory. First, we considered the transverse Ising model and the QGG. To
construct the gauge theory, we defined the gauge transformation by
rotational operator on the Hilbert space. It is essential that
interaction of these models is written in term of one or two components
of spin operators. If a system has Heisenberg-type interactions, we can
not define a gauge transformation which satisfies the commutation
rule.

Using the gauge theory, we obtained mainly two results. One is the
identity for gauge invariant operators. The FM limit state and the
classical equilibrium state on the NL provide the same expectation value
for gauge invariant operators. We note that this result remains valid
when we introduce time evolution following the Schr\"{o}dinger
equation. This result has already been pointed out for classical spin
glasses with stochastic dynamics. \cite{O1,O2} We have shown that the
same applies to quantum spin systems.

The other result is a set of inequalities for the correlation
function. These inequalities show that the correlation function for the
quantum model never exceeds the classical counterpart on the NL. The
corresponding classical system is determined by a transformation rule
for probability distribution.  As a result, the order parameter is
smaller than the square of the classical one. Therefore the
FM phase (or the KT phase) should lie within the
corresponding classical one. This is natural intuitively since quantum
effects reduce ordering tendency, but to prove it rigorously is a
different and quite a non-trivial problem. Moreover, these results are
valid even if the system is in the ground state. Thus FM long range
order vanishes in the two-dimensional QGG although the ground state of the
pure quantum $XY$ model has FM order. 

Next, we determined the phase diagram for the quantum $XY$ Mattis
model. This model is not a real SG because of lack of frustration. It is
interesting, nevertheless, that both quantum and classical phase
transitions occur in a single system, which may serve as a starting
point for investigations of more realistic quantum spin glasses with
frustration.

\section*{Acknowledgment}
This work was supported by the Grant-in-Aid for
Scientific Research on Priority Area ``Statistical-Mechanical Approach
to Probabilistic Information Processing'' by the MEXT.

\end{document}